\newif\ifprd
\prdfalse
\ifprd
\documentclass[%
aps,%
prd,%
floatfix,
eqsecnum,%
twocolumn,%
nofootinbib%
]{revtex4}
\else
\documentclass[12pt]{article}
\fi

\usepackage{ifthen}
\usepackage{calc}
\usepackage{amsmath}
\usepackage{amssymb}
\usepackage{pict2e} 
\usepackage{graphicx} 

\newcommand{\ourtitle}{Running with Triplets: How Slepton Masses Change With Doubly-Charged Higgses}
\newcommand{\paperauth}{N. Setzer and S. Spinner}
\ifprd{}
\else
\title{\ourtitle}
\author{\paperauth}	
\fi

\newcommand{\deriv}[2]{   \frac{ d {#1}}{ d {#2} }  }

\newcommand{\abs}[1]{ \left| \, {#1} \right| }

\DeclareMathOperator{\sgn}{sgn}

\newcommand{\MP}{M_{\text{Pl}}} 		
\newcommand{\MSUSY}{M_\text{SUSY}}	
\newcommand{\Mmess}{M_\text{mess}}

\newcommand{\half}{\frac{1}{2}}

\newcommand{\fourth}{\frac{1}{4}}

\DeclareMathOperator{\Tr}{Tr}

\newcommand{\inp}[2][0cm]{ \mathopen{}\left( #2 \parbox[h][#1]{0cm}{} \right) }
\newcommand{\inb}[2][0cm]{ \mathopen{}\left[ #2 \parbox[h][#1]{0cm}{} \right] }

\newcommand{\inap}[2][0cm]{ \mathopen{}\left< {#2} \parbox[h][#1]{0cm}{} \right> }

\newcommand{\pfrac}[2]{ \mathopen{}\left( \frac{#1}{#2} \right) }

\newcommand{\eq}[1]{Eq.~\eqref{Eq:#1}}

\newcommand{\fig}[1]{Figure~\ref{Fig:#1}}
\newcommand{\Sec}[1]{Section~\ref{Sec:#1}}
\newcommand{\tbl}[1]{Table~\ref{Table:#1}}

\newcommand{\E}[1]{\times 10^{#1}}
\newcommand{\vev}[1]{\inap{#1}} 

\newcommand{\MSSMDC}{MSSM$+$DC}
\newcommand{\ourmodel}{TESSM}
\newcommand{\G}[1]{\ensuremath{G_{#1}}}
\newcommand{\Wtriplets}{W_{\text{T}}}
\newcommand{\WMSSMDC}{W_{\text{\MSSMDC}}}

\begin{document}             

\ifprd
\else
\maketitle
\fi

\begin{abstract}
We examine the slepton masses of SUSYLR models and how they change due the presence of light-doubly charged higgs bosons.  We discover that the measurement of the slepton masses could bound and even predict the value of the third generation Yukawa coupling of leptons to the $SU(2)_R$ Triplets.  We also consider the unification prospects for this model with the addition of left-handed, $B - L = 0$ triplets---a model we call the Triplet Extended Supersymmetric Standard Model (\ourmodel).  Finally, we discuss the changes in the slepton masses due to the presence of the $SU(2)_L$ triplets.
\end{abstract}

\section{Introduction}
\label{Sec:Intro}

When the Large Hadron Collider (LHC) comes online in a couple of years, it is widely believed that it will discover a great deal of new physics.  This notion is well motivated---the Standard Model's Higgs boson must remain light for proper electroweak symmetry breaking and therefore the next new scale of physics must be around a TeV.  One theory for keeping the Higgs light is supersymmetry (SUSY), which is a symmetry between bosons and fermions\cite{SUSYprimer}.  Apart from just stabilizing the Higgs mass, the Minimal Supersymmetric Standard Model (MSSM) gives gauge coupling unification and contains a viable dark matter candidate---so there is a strong reason to think that it accurately describes nature around the TeV Scale.

However, while the MSSM is so appealing, it is the minimal extension of the Standard Model and therefore doesn't naturally contain right-handed neutrinos.  Since a $\nu_R$ is required for the understanding of the neutrino oscillations, it may be added as a singlet field to the model.  Yet this brings along with it a naturalness problem: the coupling of the neutrino to the Higgs-boson would give it a mass $m_D$ on the order of the known quarks and leptons---a value much too large to fit the experimentally measured oscillation data.

The solution to this quandary is to give $v_R$ a large majorana mass and thus use the seesaw mechanism\cite{seesaw} to get a small $\nu_L$ mass.  The most appealing way to implement this mechanism is to extend the gauge group from $SU(3)^c \times SU(2)_L \times U(1)_Y$ to $SU(3)^c \times SU(2)_L \times SU(2)_R \times U(1)_{B - L}$ (\G{3221}).  The extension of the gauge group adds more than just a natural reason for the right-handed neutrino: it also explains why $M_R \ll \MP$\cite{Journal:Phys.Rev.Lett.44:1316.1980:B-L.Gauged} and even allows this mass to be predicted\cite{Journal:Phys.Rev.D.73:075001.2006:PredictMR}.  Models utilizing $\G{3221}$ have been considered before\cite{susylr,muon,Journal:Phys.Rev.D.65:0116005.2002:strongCP.SUSYphase.doublets,Journal:Phys.Rev.D.60:095004.1999:upDownUnification.triplets}, and it is popular to use $SU(2)_R$ triplets to achieve the seesaw mechanism.  An additional attractive feature is that these models automatically conserve $R$-parity which allows a natural dark matter candidate.

An interesting artifact of including right-handed triplets is that in minimal SUSYLR models they result in light doubly-charged particles\cite{Journal:Phys.Rev.D.73:075001.2006:PredictMR,SUSYLRDC}.  This is due to the expanded global symmetry of the Langrangian meaning they only receive mass from non-renormalizable terms\cite{DCPhenom,biswa}.  Since the doubly-charged Higgses can survive to the TeV scale, they influence the renormalization group equations (RGEs).

Specifically, the right-handed triplets must couple to the leptons through the term $f_c L^{c T} \tau_2 \Delta^c L^c$ to give the large Majorana mass to the right-handed neutrinos.  This coupling then forces the doubly-charged particles to couple to the sleptons. Since these Higgses survive to the TeV scale, they alter the slepton RGEs and hence their masses\cite{Journal:Phys.Rev.D.59:015018.1999:DCPhenom}.  

The slepton mass running is highly dependent on $f_c$, and we will demonstrate in \Sec{GMSBDC} that one may bound $f_c$ by limits on the stau mass.  In fact, one can do better than bound $f_c$: a measurement of a right-handed selectron mass in excess of the MSSM's result, combined with a measurement of the $\tilde\tau_1$ mass, would yield a value for the third generation $f_c$.  We think that this is an important result to emphasize since probing the TeV scale slepton masses will then yield an indication of the physics roughly $8$ orders of magnitude higher.

The idea that the light doubly-charged Higgs bosons change the RGEs also compels one to ask if unification is preserved, which we address in \Sec{Unification}.  The authors of \cite{Journal:Phys.Rev.D.59:015018.1999:DCPhenom} considered this and resorted to adding more Higgs doublets.  We adopt a different tactic and use $B - L = 0$ triplets.  The reason for our approach is that the doubly-charged Higgses alter the RGEs for hypercharge but not the left-handed coupling.  Thus $\alpha_1$ runs quickly while $\alpha_2$ runs slowly, and it is necessary to include particles that will change the left-handed coupling's running while not affecting the hypercharge running.  

Further complications arise in unification from requiring right-handed coupling remain perturbative.  We find in \Sec{Unification} that $\alpha_R$ will only remain perturbative up until about $10^{12}$ GeV.  Due to this, we focus our discussion of unification on Gauge-Mediated SUSY breaking scenarios.

\section{Slepton Masses with Light Doubly Charged Higgses}
\label{Sec:GMSBDC}

In this section we consider a minimal extension of the Minimal Supersymmetric Standard Model (MSSM) to include doubly-charged Higgs superfields (DC).  We dub the resulting model the \MSSMDC.  From the view point of the MSSM, the doubly-charged particles are simply singlets of all but the hypercharge group with $Y = 4$, where $Q_{EM} = I_{3L} + \frac{Y}{2}$.  

Since this effective theory is a general low energy result of well motivated high energy theories, it would be very insightful to investigate its low energy properties.  We examine these details here, adding to the work done in \cite{Journal:Phys.Rev.D.59:015018.1999:DCPhenom}.  We also use this section as a springboard into the \Sec{Unification}, which will have similar phenomenology.

In order to facilitate this investigation, we begin with the appropriate expressions defining this model.  The DC are denoted as $\Delta^{--}$ and $\bar\Delta^{++}$ in the following superpotential and corresponding SUSY breaking potential:

\begin{equation}
\label{Eq:MSSM+DC}
\begin{aligned}[b]
\WMSSMDC	& =	u^c y_u Q H_u
	  	- d^c y_d Q H_d
	  	- e^c y_e L H_d
		+ e^{c} f_c e^c \Delta^{--}
	\\
	&	\quad {}
			+ \mu H_u H_d + \mu_\Delta \Delta^{--} \bar \Delta^{++}
\end{aligned}
\end{equation}
\begin{align}
\notag
\label{Eq:V non SUSY}
V_{\text{soft}}
	& =	\inp
			{
				\tilde u^c a_u \tilde Q H_u
				- \tilde d^c a_d \tilde Q H_d
				- \tilde e^c a_e \tilde L H_d
				+ \tilde e^c a_c \tilde e^c \Delta^{--}
				+ \text{c.c.}
			}
		\\ \notag
		& \quad {}
			+ m_{\Delta}^2 \abs{\Delta^{--}}^2
		 	+ m_{\bar{\Delta}}^2 \abs{\bar \Delta^{--}}^2
			+ m_{H_u}^2 \abs{H_u}^2
			+ m_{H_d}^2 \abs{H_d}^2
		\\ \notag
		& \quad {}
			+ \tilde Q^\dagger m_Q^2 \tilde Q
			+ \tilde L^\dagger m_L^2 \tilde L 
			+ \tilde u^c m_{u^c}^2 \tilde u^{c \dagger}
			+ \tilde d^c m_{d^c}^2 \tilde d^{c \dagger}
			+ \tilde e^c m_{e^c}^2 \tilde e^{c \dagger}
		\\
		& \quad {}
			+ \inp{b H_u H_d + b_\Delta \Delta^{++} \Delta^{--} + \text{c.c.}}
\end{align}%
Where, as usual, generational, color and isospin indices have been suppressed.

We choose to explore the mass spectrum in the gauge-mediated SUSY breaking (GMSB) scenario (for a review see \cite{SUSYprimer,Topic:GMSB}) which conjectures that SUSY breaking is communicated from some hidden sector to the matter sector via ordinary gauge interactions.  One usually assumes that these messenger fields form $N_5$ copies of complete $5 + \bar{5}$ representations of $SU(5)$ at a scale $\Mmess$, the mass of the messenger particles.  Gaugino and scalar masses at $\Mmess$ are then proportional to $\Lambda =  \frac{\inap{F}}{\Mmess}$, where $\inap{F}$ is a measure of SUSY breaking.  The benefits of GMSB scenarios are: i) their predictability (five parameters: $\tan{\beta}$, $\sgn{\mu}$, and the three just mentioned) and ii) their lack of potentially dangerous flavor violating terms.  Furthermore, in this scenario the lightest supersymmetric particle (LSP) is the gravitino/goldstino which places phenomenological importance on the next-to LSP, the NLSP---usually the lightest neutralino or stau.

We obtain our results in both models by running gauge and yukawa couplings from the scale corresponding to the mass of the $Z$ ($M_Z$) up to the messenger scale with gaugino and scalars masses based on GMSB boundary conditions.  We then run down from $\Mmess$ to $\MSUSY \sim 1$ TeV, where the tree level minimization conditions are used to solve for $b$ and $\mu$.  These minimization conditions are the same in both models.  We do all running at the one-loop level.  While more rigorous schemes exist\cite{Topic:RGEs}, our interests are in the sleptons, the 2 lightest neutralinos, and the lightest chargino masses which are adequately reproduced in this scheme.  We compared our MSSM values to ISAJET\cite{isajet} and the difference was at most $3 \%$.  Furthermore, these errors mostly cancel as our interest lies in the relative differences between the MSSM and \MSSMDC{}.

Motivated by the small yukawa values for the first and second generation quarks and leptons, the yukawa $3 \times 3$ matrix, in generation space, can be replaced by a scalar coupling for the third generation only.  While this is a common practice in the MSSM, it also works for the new yukawa coupling $f_c$ based on results from muonium oscillations and flavor violating decays\cite{Topic:muonium.osc,PDGBook} (which constrain all but the $\tau \tau$ component\cite{Journal:Phys.Rev.D.59:015018.1999:DCPhenom}).

The boundary conditions for GMSB\cite{SUSYprimer, Topic:GMSB} are:
\begin{align}
\label{Eq:GauginoBC}
	M_a 		& = \frac{\alpha_a \Lambda}{4 \pi}					
	\\
\label{Eq:ScalarBC}
	m_\phi^2		& = 2 \Lambda \inb{\inp{\frac{\alpha_a}{4 \pi}}^2 C_a^\phi}		& (a = 1,2,3)
	\\
\label{Eq:ABC}
	a_i 		& \sim 0 								& (i = u,d,e,c)
\end{align}
for gauginos, scalars, and trilinear a terms respectively (we have used the traditional normalization $g_1 = \sqrt{\frac{5}{3}} g^\prime$, and $C_a^\phi$ is the quadratic Casimir invariant).
 
The RGEs for a general SUSYLR can be found in \cite{Journal:Phys.Rev.D.71:115010.2005:SUSYLR.RGEs}, and we utilize those (with appropriate changes).  Upon minimal investigation of the RGEs, it becomes clear that $\alpha_1 = \frac{g_1^2}{4 \pi}$ will get a large contribution due to the DC.  This translates into a larger value at $\Mmess$, and hence larger mass boundary conditions for $M_1$ and the right-handed scalar masses $m_{\tilde \tau^c}$ and $m_{\tilde e^c}$ (compared to the MSSM values).  However, $M_1$ will decrease quickly as it is evolved to $\MSUSY$, and will have a value comparable to that in the MSSM; the scalar masses will actually increase.  This will not be true for the soft-breaking mass of the stau if $f_c$ is large, as this will cause the mass to decrease.  Below we list the relevant RGEs and boundary conditions:
\begin{equation}
\label{Eq:alpha1Running}
  \deriv{\alpha_1^{-1}}{t} = -\frac{3}{5} \, \frac{19}{2 \pi}
\end{equation}
\begin{align}
\label{Eq:mtaucRGE}
\notag
	16 \pi^2 \deriv{m_{\tilde \tau^c}^2}{t}
	&	=
			4 \abs{y_\tau}^2 \inp{m_{\tilde \tau^c}^2 + m_{H_d}^2 + m_{L}^2}
			+	8 \abs{f_c}^2 \inp{2 m_{\tilde \tau^c}^2 + m_{\Delta}^2}
	\\
	&	\quad {}
			+ 4 \abs{a_\tau}^2
			+ 8 \abs{a_c}^2
			- 4 \pi \inp{\frac{24}{5} \, \alpha_1 \abs{M_1}^2}
\end{align}
\begin{align}
\label{Eq:mecRGE}
	\deriv{m_{\tilde e^c}^2}{t}
	&	=
			-\frac{1}{4 \pi} \inp{\frac{24}{5} \, \alpha_1 \abs{M_1}^2}
\end{align}
along with the boundary conditions at $\MSUSY$ based on \eq{ScalarBC}:
\begin{align}
\label{Eq:mecBC}
	m_{\tilde e^c \! \!, \; \tilde \tau^c}^2
	&	=
			2 \Lambda^2 \inb{\frac{3}{5} \inp{\frac{\alpha_1}{4 \pi}}^2}
\end{align}
where $t = \ln{\frac{Q}{Q_0}}$ with $Q$ the RG scale.

The relevant mass formulae, which are the same in the MSSM and the extension considered here are:
\begin{multline}
\label{Eq:stauMass}
m_{\tilde \tau}^2
	= \\
	\begin{pmatrix}
		m_{\tilde L 3}^2 + y_\tau^2 v_d^2
		+ \frac{\pi}{2} \inp{\frac{3}{5}\alpha_1 - \alpha_2} \inp{v_d^2 - v_u^2}
		&
		\frac{1}{\sqrt2} \inp{a_\tau v_d - v_u \mu y_\tau}
	\\
		\frac{1}{\sqrt2} \inp{a_\tau v_d - v_u \mu y_\tau}
		&
		m_{\tilde \tau^c}^2 + y_\tau^2 v_d^2 - \pi \alpha_2 \inp{v_d^2 - v_u^2}
	\end{pmatrix}
\end{multline}
\begin{align}
\label{Eq:selectronRMass}
m_{\tilde e_R}^2
	&	=
	m_{\tilde e^c}^2 - \frac{3}{5} \, \pi \alpha_1 \inp{v_d^2 - v_u^2}
\end{align}

Where $m_{\tilde L 3}$ is the soft mass for the third generation slepton isospin doublet, $m_{\tilde \tau^c}$ is the soft mass for the third generation slepton isospin singlet, $v_u$ is the vev of the up-type higgs and $v_d$ is the vev of the down-type higgs.  The lighter eigenstate of \eq{stauMass} is typically called $\tilde \tau_1$, the heavier $\tilde \tau_2$.  Mass expressions for the remaining sleptons, charginos and neutralinos can be found in \cite{SUSYprimer}.  

For the standard GMSB parameters we choose the Snowmass point SPS8: $\Lambda = 100$ TeV, $\tan{\beta} = 15$, $N_5 = 1$, $\Mmess = 200$ TeV, $\sgn{\mu} = +1$ \cite{Topic:SPS}, and $Q = 1$ TeV; $Q$ being the scale at which the masses are quoted.  Furthermore, the extended model contains the additional parameters $f_c$, $\mu_{\Delta}$, and $b_{\Delta}$.  The boundary condition $b_{\Delta} = 0$ is used at $\Mmess$, and $\mu_{\Delta}$ does not have much affect on the masses of interest here.  With this in mind, we first present sparticle masses (in GeV) at two different $f_c(M_Z)$ boundary condition values, but a constant $\mu_\Delta = 800$ (\tbl{SparticleMasses}).  The table confirms significant mass differences as qualitatively discussed earlier.
\begin{table}[ht]
\begin{center}
\begin{tabular}{|c|c|c|c|c|c|}
	\hline\hline
	Sparticle		& MSSM	& MSSM+DC	& Percent 	& \MSSMDC 	&	Percent			\\
				&	& $f_c = 0.1$	& Difference	& $f_c = 0.6$	&	Difference		\\
	\hline
	$\tilde \tau_1$		& $163$	& $183$		& $12 \%$	& $118$		& $28 \%$			\\
	$\tilde \tau_2$		& $369$	& $371$		& $1 \%$		& $371$		& $1 \%$				\\
	$\tilde e_R$		& $171$	& $191$		& $12 \%$	& $191$		& $12 \%$			\\
	$\tilde e_L$		& $367$	& $369$		& $1 \%$		& $369$		& $1 \%$				\\
	$\tilde \nu_\tau$	& $358$	& $360$		& $1 \%$		& $360$		& $1 \%$				\\
	$\tilde \nu_e$		& $358$	& $361$		& $1 \%$		& $361$		& $1 \%$				\\
	$\tilde\chi_1^0$		& $132$	& $128$		& $3 \%$		& $128$		& $3 \%$				\\
	$\tilde\chi_2^0$		& $264$	& $259$		& $2 \%$		& $259$		& $2 \%$				\\
	$\tilde\chi_1^+$		& $263$	& $258$		& $2 \%$		& $258$		& $2 \%$				\\
	\hline\hline
\end{tabular}
\end{center}
\caption{Sparticle masses for $\Lambda = 100$ TeV, $\tan{\beta} = 15$, $N_5 = 1$, $\Mmess = 200$ TeV and $\sgn{\mu} = +1$, $f_c = 0.1$ and $0.6$ and $\mu_\Delta = 800$ GeV.  These masses are reported in GeVs at $\MSUSY$.  Percent differences are included for the purpose of easy comparison.}
\label{Table:SparticleMasses}
\end{table}
\begin{figure}[ht]
	\begin{center}
		\begin{picture}(0,0)
		\put(5,240){$m_{\tilde\tau_1}$ (GeV)}
		\put(365,5){$f_c$}
		\put(307,211){$\tan \beta = 5$}
		\put(307,190){$\tan \beta = 15$}
		\put(307,175){$\tan \beta = 25$}
		\end{picture}
		\includegraphics[scale=0.6]{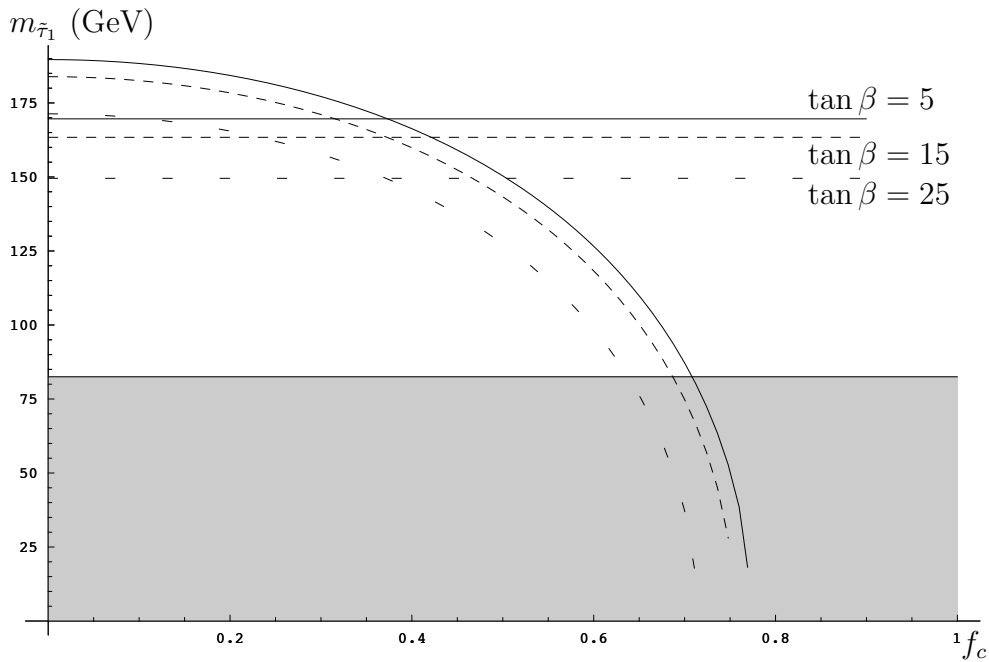}
	\end{center}
	\caption{The lightest stau mass as a function of $f_c$ in the MSSM (straight lines) and \MSSMDC{} (curves).  The shaded region is excluded by LEP II.  The graph clearly demonstrates that for a given $\tan \beta$ and $\Lambda$ there is an upper bound on $f_c$.}
	\label{Fig:StauMass}
\end{figure}
\begin{figure}[ht]
	\begin{center}
		\begin{picture}(0,0)
		\put(5,245){$\tan\beta$}
		\put(365,5){$f_c$}
		\put(120,80){$\chi_1^0$ NLSP}
		\put(240,200){$\tilde\tau_1$ NLSP}
		\end{picture}
		\includegraphics[scale=0.6]{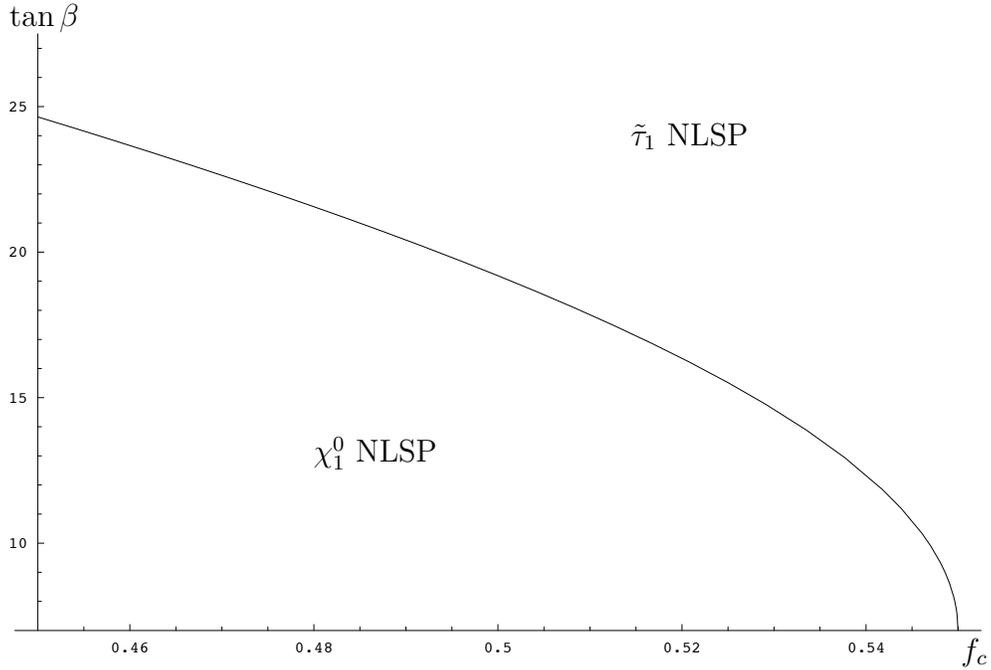}
	\end{center}
	\caption{
	This plot shows the dividing line between a neutralino LSP and a stau LSP as a function of $\tan \beta$ and $f_c$ for $\Lambda=100$ TeV.  Notice that larger values of $f_c$ favor a $\tilde \tau_1$ LSP.
	}
	\label{Fig:NLSP:tanBeta}
\end{figure}
\begin{figure}[ht]
	\begin{center}
		\begin{picture}(0,0)
		\put(110,240){$\Lambda$ (GeV)}
		\put(360,5){$f_c$}
		\put(160,130){$\chi_1^0$ NLSP}
		\put(270,70){$\tilde\tau_1$ NLSP}
		\end{picture}
		\includegraphics[scale=0.6]{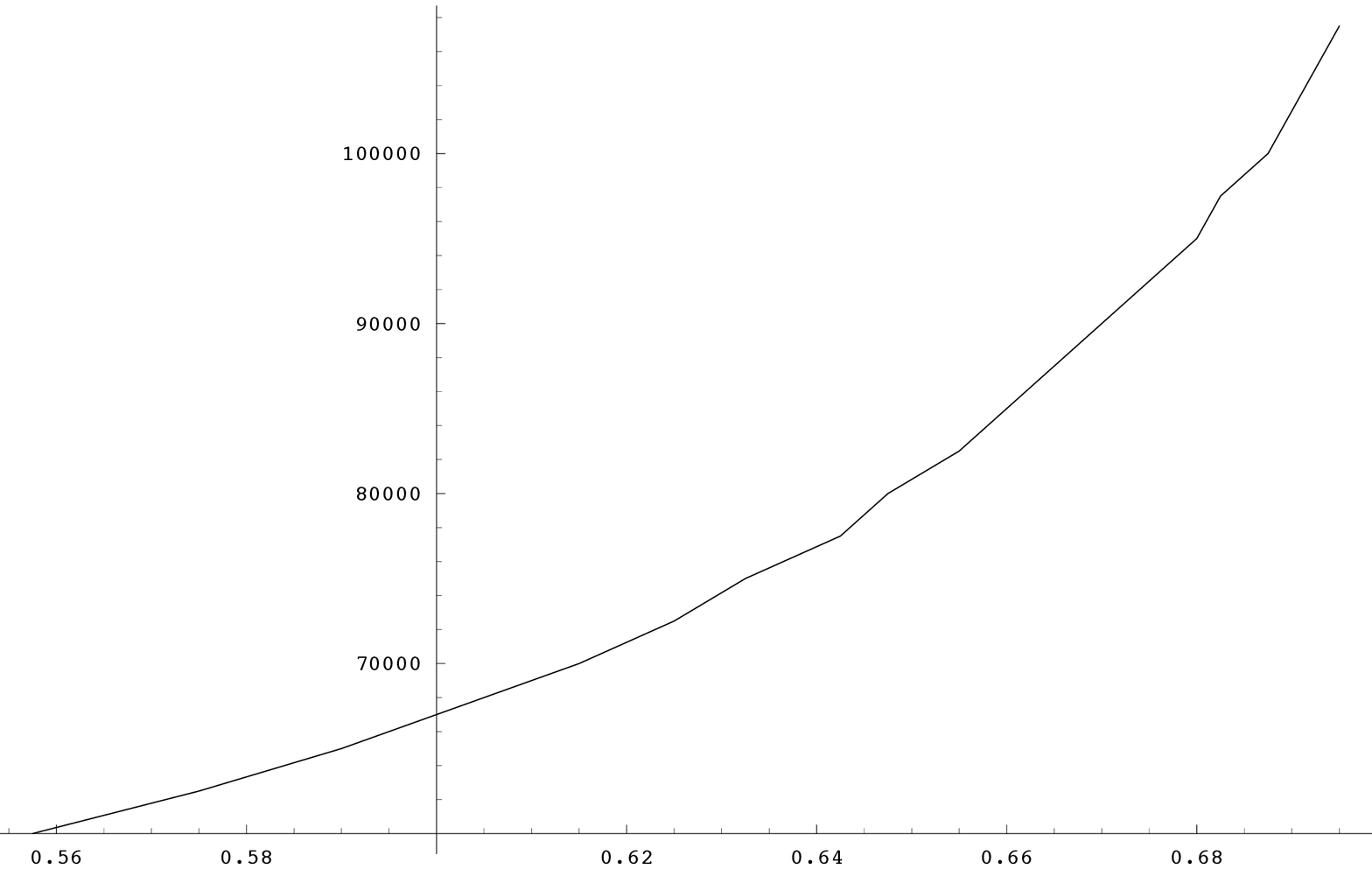}
	\end{center}
	\caption{
	This plot demonstrates the regions defining a neutralino LSP and a stau LSP as a function of $\Lambda$ and $f_c$ for $\tan \beta=15$.  A $\tilde \tau_1$ LSP occurs for higher $f_c$ values.
	}
	\label{Fig:NLSP:Lambda}
\end{figure}

To further illuminate the dependence of $m_{\tilde \tau_1}$ on $f_c$, we include \fig{StauMass}, which shows mass values at different $\tan{\beta}$ values.  The straight lines are the MSSM values and the curves with the correpsonding pattern are the matching \MSSMDC{} values.  The lower bound is from LEP II searches for NLSP staus \cite{Abdallah:2002rd} and would serve as a limit on $f_c$ given a value of $\tan{\beta}$ and $\Lambda$.

According to \fig{StauMass}, for $f_c \sim 0.5$ the stau mass drops below the neutralino mass indicating a transition from neutralino NLSP to stau.  Such a scenario is possible in the MSSM for larger values of $\tan{\beta}$ and low values of $N_5$ indicating that if an NLSP stau is discovered outside of this range, it would hint at the validity of this model---an exciting possibility.  \fig{NLSP:Lambda} shows which regions of the $\Lambda$--$f_c$ plane produce a stau NLSP and the same is done in \fig{NLSP:tanBeta} but on a $\tan{\beta}$--$f_c$ plane.

Assuming that the DCs are hard to detect, an additional indicator for this model would consist of a measurement of $\tilde e_R$ mass that is heavier than expected.  Such a measurement will be possible at a future linear collider such as the ILC but would depend on cascade chains at the LHC.  As long as $f_c$ is not at a value such that the $\tilde \tau_1$ mass is at its MSSM value, the mass of $\tilde \tau_1$ could also hint at the presence of light DCs.  Either way, measuring the mass of $\tilde \tau_1$ will yield a value for $f_c$, a parameter which has implications in the neutrino sector.

Detection of the DCs themselves would be a smoking gun for this model.  Pair-production is possible either at the LHC (through quark annihilation) or at the ILC.  Each boson would then decay into two like-signed taus at one vertex and two like-sign taus of opposite charge at the other vertex.  Background for this four tau signal should be manageable\cite{Journal:Phys.Rev.D.59:015018.1999:DCPhenom}.  Detection of the corresponding higgsinos would also be possible through the same processes; however, these would decay into a tau and a stau.

General GMSB phenomenology, which is applicable in this model, can be broken down into cases based on $\inap{F}$.  For small values, $100$ TeV or less, the NLSP will decay inside the detector.  Between $100$ to $1000$ TeV, it will decay in the detector but with a displaced vertex---which would yield information about $\inap{F}$.  If $\inap{F}$ is greater than $1000$ TeV, the NLSP decays outside the detector and therefore acts as the LSP, but only from a detector viewpoint and not from a cosmological one \cite{SUSYprimer}.  If the stau is the NLSP, this scenario will yield ionized tracks---a distinct signal of a long-lived charged particle.  LEP II has lower bounds for this at about $81$ GeV\cite{Barate:1998zp}.

Decays of an NLSP stau will produce missing energy${} + \tau$.  Staus will most likely be a product of pair produced heavier sleptons decaying to $\tilde \ell \rightarrow \tilde \tau_1 + \tau + \ell$ hence giving rise to a final signal of four taus plus missing energy.  At the LHC this would be accompanied by jets.  Although co-NLSP right-handed sleptons are also a possibility in some parts of the parameter space, in the case of interest here---larger $f_c$ value---this will not occur.

For further discussions in this model see \cite{Journal:Phys.Rev.D.59:015018.1999:DCPhenom} and in general GMSB see \cite{Topic:GMSB, SUSYprimer}.  For small values of $f_c$, the neutralino---which is mostly bino---is the NLSP and its most likely decay mode is to a photon and gravitino.  At a linear collider, decays of $\chi_1^0$ inside the dector will yield two photons plus missing energy with removable standard model (SM) backgrounds.  The photons in this case make detection of SUSY particles easier than the neutralino LSP case, which only produces missing energy.  Detection at the LHC could proceed through slepton pair production via valence quark interactions.  These would eventually decay into 2 photons $+$ missing energy $+$ leptons and jets---although a more dominant mode would be through a mixed or pair production of squark and gluino and decaying to 2 photons $+$ missing energy $+$ jets.

\section{Unification}
\label{Sec:Unification}

The gauge coupling unification of models with DC Higgses has been discussed \cite{Journal:Phys.Rev.D.59:015018.1999:DCPhenom}, and it was pointed out that the couplings may be chosen to unify at around $10^{12}$ GeV; however, when \cite{Journal:Phys.Rev.D.59:015018.1999:DCPhenom} considered unification, the authors chose to have two additional Higgs doublets at $10$ TeV.  We will present an alternative solution that maintains the usual two Higgs doublets at low scales and requires the additional particle content to have masses at the TeV scale.  

To motivate our solution, we first note that the DC Higgs bosons only affect hypercharge---causing a drastic increase in the running.  Since we wish to unify to \G{3221} this presents a major problem: if the left- and right-handed couplings run the same way, then both will run too slowly and force $\bar{\alpha}_{BL}^{-1}$ to be non-perturabtive or even less than zero at the right-handed scale.  This is because the hypercharge will run so quickly that it will be very close to the left-handed coupling at unification, which implies it is close to the right-handed coupling, and since the $B-L$ coupling is the difference, it is close to, or less than, zero.

\begin{figure}[ht]
\ifprd
\else
\begin{picture}(0,0)
\put(46,217){$\alpha_1^{-1}$}
\put(46,178){$\alpha_2^{-1}$}
\put(46,70){$\alpha_3^{-1}$}
\put(270,165){$\alpha_{3L}^{-1} = \alpha_{2L}^{-1} = \alpha_{1L}^{-1}$}
\put(270,13){$\alpha_{3R}^{-1} = \alpha_{2R}^{-1} = \alpha_{1R}^{-1}$}
\put(150,33){$t_{\text{mess}}$}
\put(367,33){$t$}
\end{picture}
\includegraphics[scale=0.6]{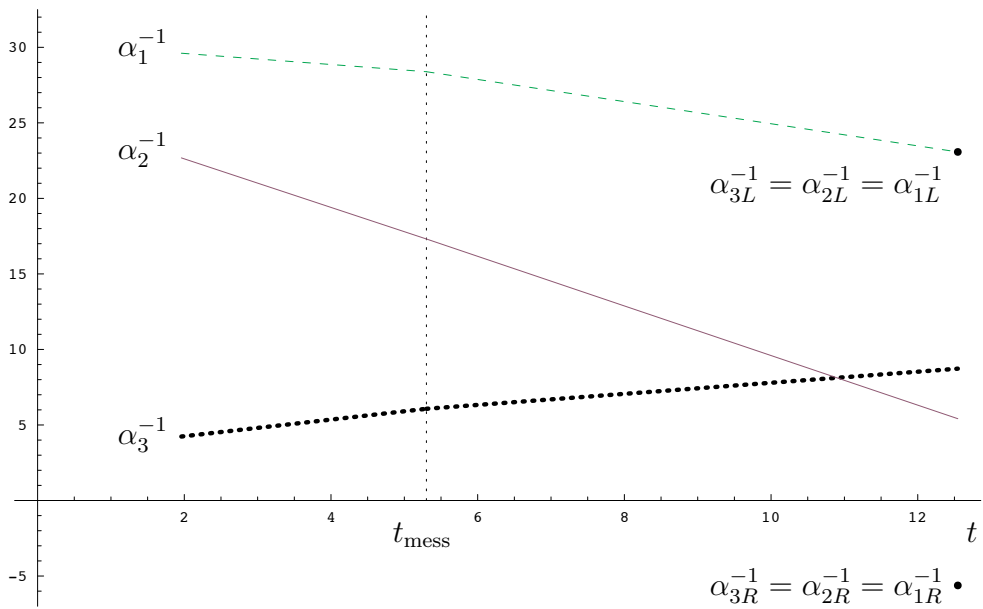}
\fi%
\caption{
The running of the gauge couplings for the \MSSMDC.  The horizontal axis is $t = \log_{10}(\Lambda/\text{GeV})$.  The higher scale theory is assumed to be $SU(5)_L \times SU(5)_R$ since $SO(10)$ is not a viable option.  Notice that while ``unification" can be achieved, $\alpha_{2R}^{-1}$ is negative there, and thus non-perturbative well before unification. The label $t_{\text{mess}} = \log_{10}\pfrac{M_\text{mess}}{\text{GeV}}$ indicates the scale where the Messenger particles become important.
}
\label{Fig:Gauge Coupling MSSM+DC}
\end{figure}

One way to evade this problem is to abandon the parity symmetry---and with it Gravity Mediated SUSY breaking models.  The best candidate is then Gauge Mediated SUSY breaking (GMSB), which will allow the right-handed and left-handed couplings to run differently.  While an improvement (see \fig{Gauge Coupling MSSM+DC}), the two couplings do not diverge quickly enough and so the problem remains.

Since it is the ``slowness" of the left-handed coupling that is the issue, it becomes necessary to find a means to make this coupling run faster while not influencing the other couplings.  It is also desired to make $\alpha_2$ run much faster, so a particle in a representation higher than the fundamental should be added.  We choose to add the simplest higher representation: $B-L = 0$, $SU(2)_L$ triplets.  We name this the Triplet Extended Supersymmetric Standard Model (\ourmodel), and its particle content is shown in \tbl{\ourmodel}.  The couplings can then be made to unify in two ways.

\begin{table}[ht]
\begin{center}
\begin{tabular}{|l|ccccc|}
\hline\hline
		& $SU(3)^c$	&$\times$& $SU(2)_L$	&$\times$& $U(1)_{Y}$		\\
\hline
$Q$		& 3		&	 & 2		&	 & $+\frac{1}{3}$	\\
$u^c$		& 3		&	 & 1		&	 & $-\frac{4}{3}$	\\
$d^c$		& 3		&	 & 1		&	 & $+\frac{2}{3}$	\\
$L$		& 1		&	 & 2		&	 & $-1$			\\
$e^c$		& 1		&	 & 1		&	 & $+2$			\\
$H_u$		& 1		&	 & 2		&	 & $+1$			\\
$H_d$		& 1		&	 & 2		&	 & $-1$			\\
$\Delta^{--}$	& 1		&	 & 2		&	 & $-4$			\\
$\bar{\Delta}^{++}$
		& 1		&	 & 2		&	 & $+4$			\\
$\delta_a$	& 1		&	 & 3		&	 & $0$			\\
\hline\hline
\end{tabular}
\end{center}
\caption{The particle content of the \ourmodel}
\label{Table:\ourmodel}
\end{table}

\begin{figure}[ht]
\ifprd
\else
\begin{picture}(0,0)
\put(47,230){$\alpha_1^{-1}$}
\put(47,123){$\alpha_2^{-1}$}
\put(47,44){$\alpha_3^{-1}$}
\put(338,105){$\bar{\alpha}_{BL}^{-1}$}
\put(232,40){$\alpha_{2R}^{-1}$}
\put(257,44){\vector(866,500){73}}
\put(107,0){$t_{\mu_\delta}$}
\put(156,0){$t_{\text{mess}}$}
\put(324,0){$t_{v_R}$}
\put(367,0){$t$}
\end{picture}
\includegraphics[scale=0.6]{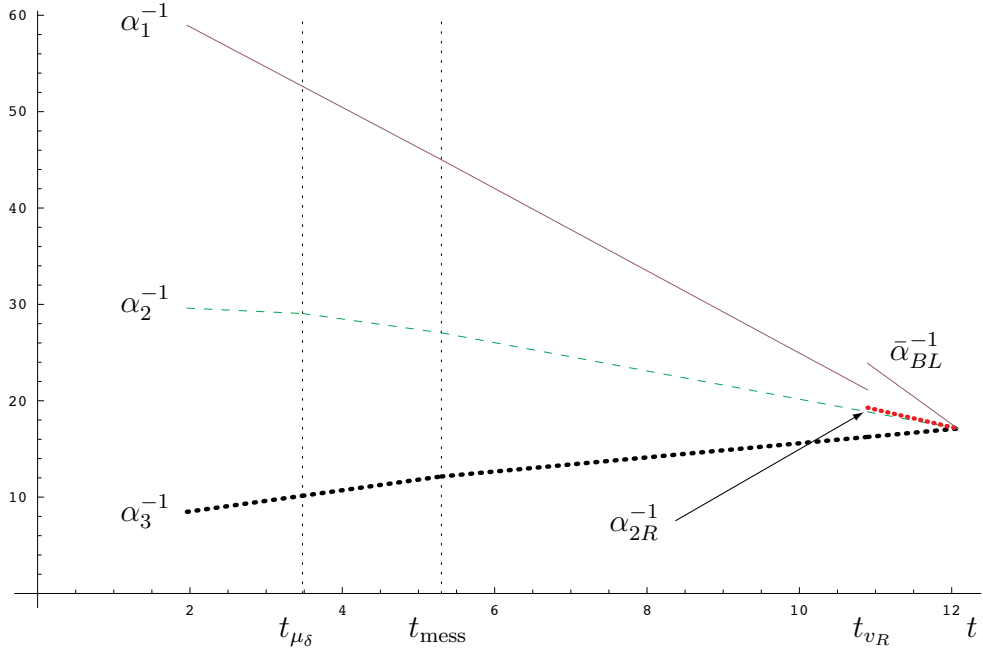}
\fi
\caption{
The running of the gauge couplings for the \MSSMDC{} and one additional $SU(2)_L$ triplet using Gauge Mediated SUSY breaking.  The horizontal axis is $t = \log_{10}(\Lambda/\text{GeV})$.  The $U(1)$ couplings have been normalized using the $SU(5)$ scheme: $\alpha_1 = \frac{5}{3} \alpha^\prime$, $\bar{\alpha}_{BL} = \frac{2}{3} \alpha_{BL}$.   The couplings unify at $t_{GUT} = 12.1$ and the right-handed scale is at $t_{v_R} = 10.9$.  The labels $t_{\mu_\delta} = \log_{10}\pfrac{\mu_\delta}{\text{GeV}}$ and $t_{\text{mess}} = \log_{10}\pfrac{M_\text{mess}}{\text{GeV}}$ indicate the respective scales where the $Y = 0$ triplets and the Messenger particles become important.
}
\label{Fig:Gauge Coupling GUT \ourmodel}
\end{figure}

The first example of unification is shown in \fig{Gauge Coupling GUT \ourmodel} where all the \G{3221} couplings unify at a scale of $M_{GUT} = 1.3 \times 10^{12}$ GeV.  This scale is far too low for $SO(10)$ (due to proton decay constraints), but any group that conserves baryon number would suffice.  To achieve this scenario it is only necessary to add one $Y = 0$ triplet, so in this sense it is the minimal model and the one on which we will focus our detailed analysis (\Sec{Our Model}).

Alternatively, the couplings may unify to $SU(5)_L \times SU(5)_R$---since we have already abandoned the parity symmetry, this group is attractive because it requires the gauge couplings to be unequal at the the unification scale\cite{Journal:Phys.Rev.D.54:5728.1996:SU5xSU5}.  Taking this as our ultimate unifying group, we may approach the problem as allowing right-handed unification at $v_R$ and then $SU(5)^2$ unification at $M_U$; however, it is quickly realized that $v_R \simeq M_U$ so it makes sense to take $v_R = M_U$.  

\begin{figure}[ht]
\ifprd
\else
\begin{picture}(0,0)
\put(25,233){$\alpha_1^{-1}$}
\put(25,180){$\alpha_2^{-1}$}
\put(25,58){$\alpha_3^{-1}$}
\put(265,183){$\alpha_{3L}^{-1} = \alpha_{2L}^{-1} = \alpha_{1L}^{-1}$}
\put(313,175){\vector(100,-132){50}}
\put(265,47){$\alpha_{3R}^{-1} = \alpha_{2R}^{-1} = \alpha_{1R}^{-1}$}
\put(62,5){$t_{\mu_\delta}$}
\put(125,5){$t_{\text{mess}}$}
\put(367,5){$t$}
\end{picture}
\includegraphics[scale=0.6]{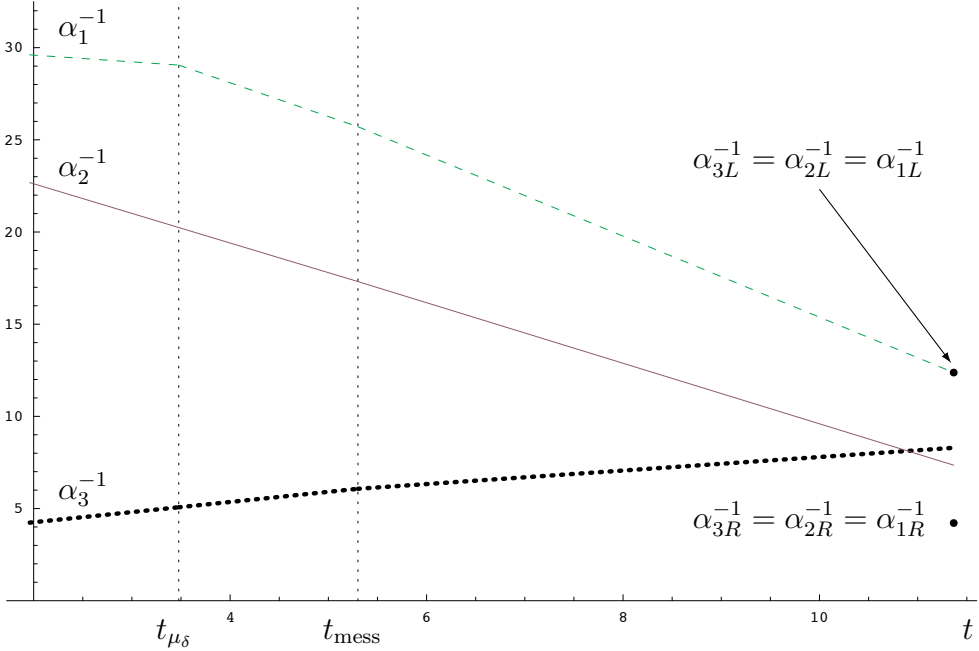}
\fi
\caption{
The running of the gauge couplings for \ourmodel{} having two $Y = 0$ triplets.  The horizontal axis is $t = \log_{10}(\Lambda/\text{GeV})$.  The theory unifies to $SU(5)_L \times SU(5)_R$ at $t_U = 11.4$.  The labels $t_{\mu_\delta} = \log_{10}\pfrac{\mu_\delta}{\text{GeV}}$ and $t_{\text{mess}} = \log_{10}\pfrac{M_\text{mess}}{\text{GeV}}$ indicate the respective scales where the $Y = 0$ triplets and the Messenger particles become important.
}
\label{Fig:Gauge Coupling \ourmodel}
\end{figure}

Given these conditions, the couplings then unify at $M_U = v_R = 2.5 \E{11}$ GeV (\fig{Gauge Coupling \ourmodel}).  To realize this unification requires the inclusion of two $B-L = 0$ triplets to the model, thus making it in some sense ``less minimal" than the previous scenario.

\section{Triplet Extended Supersymmetric Standard Model}
\label{Sec:Our Model}

The Higgs sector of \ourmodel{} has been discussed previously in \cite{Journal:Nucl.Phys.B.384:113.1992:Y=0.Triplets}, though the addition of left-handed triplets was {\it ad hoc}.  The authors of \cite{Journal:Nucl.Phys.B.384:113.1992:Y=0.Triplets} do a thorough analysis of the vacuum structure and Higgs masses; however, since they do not assume any higher scale physics, their parameters are largely unconstrained.  Our investigations show that the assumption of unification limits the parameter space to exclude the scenarios considered in \cite{Journal:Nucl.Phys.B.384:113.1992:Y=0.Triplets}.

To see the origins of these constraints, we start with the superpotential
\begin{equation}
W = \WMSSMDC + \Wtriplets
\end{equation}
where
\begin{equation}
\label{Eq:SuperW:Triplet Terms}
\Wtriplets = \mu_\delta \Tr\delta^2 + i y_\delta H_u^T \tau_2 \delta H_d
\end{equation}
and the soft breaking terms
\begin{equation}
\label{Eq:VSoft:Triplet Terms}
V_{\text{soft,T}}
	 =   m_{\delta}^2 \Tr \abs{\delta}^2
	    + \inb{ b_\delta \Tr \delta^2 + i a_\delta H_u^T \tau_2 \delta H_d + \text{h.c.} }
\end{equation}
These new terms modify the MSSM minimization conditions, but, more interestingly, add the new constraint\footnote{We assume $\vev{\delta} = v_\delta/\sqrt{2}$, $\vev{H_u^0} = v_u/\sqrt{2}$, and $\vev{H_d^0} = v_d / \sqrt{2}$; the VEVs being real.  Furthermore we take the standard $v_u = v \sin \beta$ and $v_d = v \cos \beta$.}
\begin{multline}
\label{Eq:delta min condition}
4\mu_\delta^2 + m_\delta^2 + 2 b_\delta
	+ \fourth y_\delta^2 v^2
	+ \half \frac{v^2}{v_\delta} y_\delta \mu
	- \half \frac{v^2}{v_\delta} \inp{y_\delta \mu_\delta + \half a_\delta} \sin 2\beta
 = 0
\end{multline}

Additionally, they alter the stability requirements to include
\begin{equation}
\label{Eq:delta stability}
4 \mu_\delta^2 + m_\delta^2 > \abs{2 b_\delta}
\end{equation}

Electroweak precision measurements imply that $v_\delta \ll v$, so that the terms involving $v/v_\delta$ in \eq{delta min condition} are much larger than the SUSY breaking scale.  GMSB, meanwhile, predicts that the trilinear $A$-terms are very small, and so approximately zero.  Rewriting \eq{delta min condition} keeping only the important terms (and assuming $\sin 2\beta \approx 1$, $y_\delta \approx 1$) gives
\begin{equation}
4 \mu_\delta^2 + b_\delta - \half \frac{v^2}{v_\delta} \mu_\delta + \half \frac{v^2}{v_\delta} \mu = 0
\end{equation}
The last term is large and positive, and forces $b_\delta$ to be negative with a large magnitude (since $\mu_\delta$ comes with terms of opposite sign, its contributions mostly cancel each other).  With $\abs{b_\delta}$ large, the stability condition of \eq{delta stability} requires that $\mu_\delta$ also be large (given $m_\delta^2 \sim \MSUSY^2$).  It is therefore necessary for the new $Y = 0$ triplets to be ``heavy", and our numerical analysis indicates they are around $5$ TeV.

We consider now the slepton masses.  The expressions for their masses remain the same as \MSSMDC{} except for the stau mass matrix, which is now given by
\begin{multline}
m_{\tilde \tau}^2
	= \\
	\begin{pmatrix}
		m_{\tilde L 3}^2 + y_\tau^2 v_d^2
		+ \frac{\pi}{2} \inp{\frac{3}{5}\alpha_1 - \alpha_2 } \inp{v_d^2 - v_u^2}
		&
		\frac{1}{\sqrt2} \inp{a_\tau v_d + \half y_\delta y_\tau v_u v_\delta - v_u \mu y_\tau}
	\\
		\frac{1}{\sqrt2} \inp{a_\tau v_d + \half y_\delta y_\tau v_u v_\delta - v_u \mu y_\tau}
		&
		m_{\tilde \tau^c}^2 + \abs{y_\tau}^2 v_d^2 - \pi \alpha_2 \inp{v_d^2 - v_u^2}
	\end{pmatrix}
\end{multline}

The running for these masses is slightly more complicated than in \MSSMDC, but we handle it in a similar manner.  The complication is due to the new minimization condition and two new parameters: $b_\delta$ and $\mu_\delta$.  Because of the above mentioned constraints on $\mu_\delta$ we choose it to be $5$ TeV at $\MSUSY$ and use the new minimization condition to solve for $b_\delta$.  The RGEs are derived using \cite{Martin:1993zk}.

The resulting right-handed slepton spectrum is very similar to \MSSMDC.  The left-handed sleptons, however, now get an increase to their boundary condition value which results in a higher slepton mass.  We display these masses in \tbl{CHASSMSparticleMasses} for the SPS8 point: $\Lambda = 100$ TeV, $\Mmess = 200$ TeV, $N_5 = 1$, $\tan{\beta} = 15$ and $\sgn{\mu} = +1$.  We also choose $f_c = 0.1$, $\mu_\Delta = 800$ GeV, and $y_\delta = 0.1$.  Masses are reported in GeV at $\MSUSY$, and there are no significant changes to masses with changes in $y_\delta$.  Changes with $f_c$ are as mentioned in \Sec{GMSBDC}.  The two lightest neutralinos as well as the lightest chargino do not experience any significant changes to their values.

\begin{table}[ht]
\begin{center}
\begin{tabular}{|c|c|c|c|}
	\hline\hline
	Sparticle		& MSSM	& \ourmodel	& Percent 				\\
				&	& $f_c = 0.1$	& Difference				\\
	\hline
	$\tilde \tau_1$		& $163$	& $183$		& $12 \%$				\\
	$\tilde \tau_2$		& $369$	& $385$		& $4 \%$					\\
	$\tilde e_R$		& $171$	& $191$		& $12 \%$				\\
	$\tilde e_L$		& $367$	& $381$		& $4 \%$					\\
	$\tilde \nu_\tau$	& $358$	& $375$		& $5 \%$					\\
	$\tilde \nu_e$		& $358$	& $372$		& $4 \%$					\\
	\hline\hline
\end{tabular}
\end{center}
\caption{Sparticle masses for $\Lambda = 100$ TeV, $\tan{\beta} = 15$, $N_5 = 1$, $\Mmess = 200$ TeV and $\sgn{\mu} = +1$, $f_c = 0.1$, $\mu_\Delta = 800$ GeV and $y_\delta = 0.1$ at 1 TeV in both the MSSM and \ourmodel.  Percent differences are included for the purpose of easy comparison.  All masses are in GeV.}
\label{Table:CHASSMSparticleMasses}
\end{table}

The slepton signatures at colliders are very similar to those mentioned in \Sec{GMSBDC}; however, for low values of $f_c$ all of the slepton masses will be higher than the MSSM values.  This might be misconstrued as a larger value of $\Lambda$ unless slepton masses can be compared to neutralino and chargino masses (which will be at their MSSM values).

As for the MSSM higgs sector, there will be no new radiative mass corrections \cite{Journal:Nucl.Phys.B.384:113.1992:Y=0.Triplets,Journal:Phys.Lett.B.279:92.1992:NonMinSUSY.incl.Y=0.Triplets,Journal:Int.J.Mod.Phys.A.17.4:465.2002:Y=0.Triplets.Rad.Corrections,Journal:Phys.Rev.D.71:073004.2005:Y=0.Triplets.Higgs.Decays}; however, the higgs sector is obviously expanded and there is a new vev, $\vev{\delta}$.  This vev is constrained by the $\rho$ parameter to be less than about $1.7$ GeV\cite{PDGBook}, so we take it to be around $1$ GeV.  The extended Higgs sector is composed of a neutral scalar $H_\delta^0$, a neutral pseudo-scalar $B^0$, and two singly charged scalars $H_{\delta 1}^+$ and $H_{\delta 2}^+$.  These fields will not mix very much with the MSSM fields because of the large $\mu_\delta$ value.  In \tbl{HiggsMasses} we take a quick peek at their typical tree level masses for the parameters used in the slepton table.

\begin{table}[ht]
\begin{center}
\begin{tabular}{|c|c|}
	\hline\hline
	Higgs Boson		& Mass			\\
				& (TeV)			\\
	\hline
	$H_\delta^0$		& $0.742$		\\
	$B^0$			& $14.2$			\\
	$H_{\delta 1}^+$		& $0.742$		\\
	$H_{\delta 2}^+$		& $14.2$			\\
	\hline\hline
\end{tabular}
\end{center}
\caption{Additional Higgs masses (reported at $1$ TeV) in \ourmodel for $\Lambda = 100$ TeV, $\tan{\beta} = 15$, $N_5 = 1$, $\Mmess = 200$ TeV, $\sgn{\mu} = +1$, $f_c = 0.1$, $\mu_\delta = 5$ TeV and $y_\delta = 0.1$}
\label{Table:HiggsMasses}
\end{table}

It is worth noting that there is one charged field that is degenerate with the scalar and one degenerate with the pseudo-scalar.  This relation will hold to a good extent even after radiative corrections because these fields do not significantly couple to the top sector.  Their only coupling to the top/stop is from their mixing with the MSSM Higgs sector, which is very small.  Still, the lighter fields can be paired produced via $W$ boson fusion and have electroweak-magnitude cross sections at the LHC.  If produced, they will decay into MSSM higgs fields or two electroweak bosons depending on the size of $y_\delta$.  Signatures in linear colliders for this model are discussed in \cite{Journal:Nucl.Phys.B.384:113.1992:Y=0.Triplets,Journal:Int.J.Mod.Phys.A.17.4:465.2002:Y=0.Triplets.Rad.Corrections,Journal:Phys.Rev.D.71:073004.2005:Y=0.Triplets.Higgs.Decays}.

\section{Conclusion}

We have considered an extension of the MSSM with light doubly-charged higgs bosons.  We showed that the right-handed slepton masses in this case will be significantly different in this model and verifications of these mass deviations at a linear collider will be a good signal for this model---even if the doubly charged higgses that cause these mass differences are beyond the reach of future accelerators.  In addition, the parameter space for a stau NLSP is greater than in the standard gauge mediated susy breaking scenario, implying that if the stau is found to be the NLSP (and $N_5$ and $\tan{\beta}$ are low), then this model could be an accurate description of TeV range physics.  Furthermore, measurements of the lightest stau mass, regardless of whether it is the NLSP or not will, will fix the value of $f_c$ and this has implications in neutrino physics.  

We also showed that this model has unification in two different schemes by adding left-handed triplets.  The resulting phenomenology for this \ourmodel{} model includes all of the features of the \MSSMDC{}, but is more rich with heavier left-handed sleptons and an expanded higgs sector.  The vev of the additional higgses is suppressed by the $\rho$ parameter, which leads to rigid constraints on the parameters.  This effectively forces half of the new higgses to be well outside the reach of future colliders, but potentially leaves the other half within the LHC's grasp (depending on the parameters).

\section{Acknowledgments}

We would like to thank Rabindra Mohapatra for useful conversations regarding this model.  This work is supported by the National Science Foundation grant no.~Phy-0354401.



\end{document}